\documentstyle[preprint,aps]{revtex}
\begin{document}
\draft
\title{Quantum Dynamics in Regions of Quaternionic Curvature}
\author{S. P. Brumby\cite{SPB} and G. C. Joshi\cite{JOSHI}}
\address{School of Physics, University of Melbourne, %
Research Centre for High Energy Physics,\\%
Parkville, Victoria 3052, Australia.}
\maketitle
\begin{center}
UM--P--94/24 ; OZ--94/11

hep-th/9406191
\end{center}
\begin{abstract}
The complex unit appearing in the equations of quantum mechanics is
generalised to a quaternionic structure on spacetime, leading to the
consideration of complex quantum mechanical particles whose dynamical
behaviour is governed by inhomogeneous Dirac and Schr\"{o}dinger equations.
Mixing of hyper-complex components of wavefunctions occurs through their
interaction with potentials dissipative into the extra quaternionic degrees of
freedom.
An interferometric experiment is analysed to illustrate the effect.
\end{abstract}
\pacs{PACS numbers: 	03.65.Bz, % Foundations of Quantum Mechanics
			11.30.Ly, % Other internal and higher symmetries
			03.65.Ge  % Solutions of wave equations, bound states
}
\narrowtext

\section{Introduction}
The foundations of quaternionic quantum mechanics (QQM) were laid by
Finkelstein, Jauch, Schiminovich and Speiser\cite{FJSS}.
They sought to generalise standard complex quantum
mechanics (CQM) by introducing additional geometrical concepts.  Guiding
them in their enterprise was the example of Einstein's geometrisation of
gravity.  Thus, they were led to proposing a quantum theory that was locally
identical to CQM, but with a generalised global structure requiring the
introduction of a connection on the spacetime manifold,
called the Q--connection, in order to relate complex algebras, and hence
states and measurements, at different points.  A nontrivial global structure
was described by a nonvanishing Q--curvature operator, defined to be the
commutator of the Q--covariant derivative at each point.  The theory is
complicated by ambiguity over the construction of tensor products, \cite{Nash%
,ARLH}, and questions of how complex analytic techniques (e.g.\ harmonic
analysis) are to be carried over to QQM's nontrivial bundle of complex
state-spaces over spacetime.

Given that investigations of quantum
gravity have suggested that deBroglie--wave scattering off regions of
gravitational curvature leads to effective particle creation, (e.g.\ Hawking
radiation\cite{Hawk}), we should allow the possibility that regions of
nonzero Q--curvature can act as sources and sinks of quantum probability.

More recently, \cite{Adler,LHLB,BMAD}, the zero Q--curvature (i.e.\
Q--flat) limit of the theory has been investigated.  Coupling of the
wavefunction components only occurs in the presence of quaternionic
potentials.  A common feature of these investigations is the adoption of
conventions regarding the ordering of quaternionic factors in the dynamical
equations, resulting in exponentially decaying hyper-complex components.
Thus, the Q--flat limit introduces physics that is in principle
difficult to observe.

We instead consider particle dynamics in the intermediate case termed the
``electromagnetic'' limit by \cite{FJSS}.
As in the Q--flat limit, the presence of potentials with quaternionic
elements causes mixing of the wavefunction components.

Our approach to QQM is guided by an analogy with the importance of Killing
vector fields in general relativity.  The compatibility of the metric tensor
with the covariant derivative is a very strong constraint on the connection,
simplifying the analysis of the system's consequent dynamical behaviour.
In the QQM case, imposing an analogous constraint on the Q--connection
singles out a field of unit, pure imaginary quaternions which we
identify with the $i$ of CQM and with which we generalise complex analysis.
Further, we adopt a convention
regarding the ordering of factors in the dynamical equations
which has major implications for the experimental verification of the theory.
We use Green functions to analyse an interferometric experiment in the
presence of weak quaternionic potentials.  Multiple barrier potentials are
then tractable using this technique in the weak potential limit.
Our results differ qualitatively from those of previous investigators
\cite{BMAD,Peres}.

\section{General Results}
We generalise from the CQM formalism by introducing the correspondence
\begin{equation}
\hat{E}^{CQM}\Psi = (\partial_{t}\Psi)i\hbar \;\leftrightarrow \;\;%
\hat{E}^{QQM}\Psi = (D_{t}\Psi)\eta\hbar\,,
\end{equation}
where the new (Lorentz scalar) field on spacetime $\eta (x^{\mu})$ is
defined formally by
\begin{equation}
\eta ^{\ast} = -\eta \,,\;\;\eta ^{2} = -1\,, \label{defETA}
\end{equation}
and now, because the algebra of quaternions\cite{algH} is noncommutative,
the order of the factors is crucial.

That is, the canonical 4--momentum operator of the quaternionic theory
acts on a state $\Psi $ according to the prescription
\begin{equation}
P_{\mu}\Psi = (D_{\mu}\Psi)\eta = (\partial_{\mu}\Psi %
+ \case{1}{2}[{\cal Q}_{\mu}, \Psi ])\eta\,,
\end{equation}
where $D_{\mu}$ is the Q--covariant derivative and ${\cal Q}_{j}$ is the
Q--connection.

By imposing the compatibility condition that $D_{\mu }\eta  \equiv 0$, we can
implement the programme of canonical quantisation, with $\hbar = 1$ and
$c = 1$ in the appropriate units,
\begin{eqnarray}
[X^{j} , X^{k}]=0\,,\;\;\; & [P_{j} , P_{k}]=0\,,\;\;\; & [X^{j} , P_{k}]%
= \eta \delta ^{j}_{\ k}\,.
\end{eqnarray}

The algebra of observables generated from the fundamental operators
$X^{j},\,P_{k},\,E=P_{o},\,\eta \openone $, is formally identical to the
algebra of operators of CQM.
The possibility remains that there exist operators with additional
quaternionic components. We use the symplectic decomposition of any
operator at a point,
\begin{equation}
{\cal O} = {\cal O}^{\eta } + \zeta {\cal O}^{\zeta }\,,
\end{equation}
where ${\cal O}^{\eta },\,{\cal O}^{\zeta }$ are $\eta $--complex, (i.e.\ %
a real linear combination of the unit operator, $\openone$, and the operator
$\eta \openone$). Here we have introduced the new quaternionic field
$\zeta = \zeta (x^{\mu })$ with properties
\begin{equation}
\zeta ^{2} = -1\,,\;\; \{ \eta (x^{\mu }) , \zeta (x^{\mu }) \} = 0\,.%
\label{defZETA}
\end{equation}
Note that $D_{\mu }\zeta \neq 0$ in general, and that ${\cal O}^{\zeta }$
is not necessarily zero.  The dynamical evolution of fully quaternionic
operators is expected to be complicated.

Also, there is no {\em a priori} reason why the wavefunction
(considering single particle quantum systems) has to be restricted to be
$\eta $--complex.  Instead, in the spirit of Finkelstein {\em et al.}, we
decompose the wavefunction into its natural symplectic components at each
point of spacetime
\begin{equation}
\Psi = \Psi _{\eta } + \zeta \Psi _{\zeta }\,.
\end{equation}

We interpret the probability measure associated with the first symplectic
component of the quaternionic wavefunction as corresponding to the usual
probability distribution of CQM.
Our convention regarding the order of Q--covariant differentiation and
multiplication by $\eta$ ensures that the second symplectic component of
the quaternionic wavefunction will be oscillatory in free space,
and, hence, asymptotically relevant.

In the relativistic regime, the full Q--curvature of the system is given by
$[D_{\mu } , D_{\nu }]$, where
\begin{mathletters}
\begin{equation}
[D_{\mu } , D_{\nu }]\Psi \equiv \case{1}{2}[{\cal K}_{\mu \nu },\Psi ]\,,
\end{equation}
\begin{equation}
{\cal K}_{\mu \nu } = {\cal Q}_{\nu , \mu} - {\cal Q}_{\mu , \nu}
+ \case{1}{2}[{\cal Q}_{\mu } , {\cal Q}_{\nu }].
\end{equation}
\end{mathletters}
The compatibility condition on the covariant derivative of $\eta $ implies that
$[{\cal K}_{\mu \nu }, \eta ]=0$, but in general
$[{\cal K}_{\mu \nu }, \zeta ]$ will not vanish.

This convention also allows us to treat the components of the Q--wavefunction
under EM--gauge transformation in a way formally identical to the CQM case:
\begin{mathletters}
\begin{equation}
D_{\mu }\Psi \rightarrow  D_{\mu }^{A}\Psi = D_{\mu }\Psi + e\Psi A_{\mu }%
\eta\,,\;\;P_{\mu }\Psi \rightarrow  (D_{\mu }^{A}\Psi)\eta\,, \label{GCovD}
\end{equation}
where the components of the EM--field $A_{\mu }$ are real.
Then under any EM--gauge transformation, where $\Theta(x^{\mu})$ is a real
function,
\begin{equation}
\Psi \rightarrow  \Psi ^{A} = \Psi \exp [-e\Theta(x^{\mu})\eta]\,,\;\;%
A_{\mu } \rightarrow  A_{\mu } + \Theta_{,\mu}\,.
\end{equation}
\end{mathletters}
The Aharonov--Bohm effect can be analysed in the quaternionic case in an
analogous fashion\cite{BJwip}.

The most general linear combination of the quaternion generators that
satisfies the modulus constraint on $\eta $ is
\begin{equation}
\eta (x^{\mu }) = \sin\theta \cos\phi i_{1} + \sin\theta \sin\phi i_{2} + %
\cos\theta i_{3}\,,
\end{equation}
where $\theta = \theta (x^{\mu })$ and $\phi = \phi (x^{\mu })$ are real
functions on spacetime.
Then the most general $\zeta $ field that anticommutes with $\eta $ and has
unit modulus is
\begin{equation}
\zeta (x^{\mu }) = \cos\theta \cos\phi i_{1} + \cos\theta \sin\phi i_{2}%
- \sin\theta i_{3}\,.
\end{equation}
The third generator of the quaternionic algebra at $x^{\mu}$ is
\begin{equation}
\xi (x^{\mu }) \equiv  \case{1}{2}[\eta ,\zeta ] = -\sin\phi i_{1} + \cos%
\phi i_{2} + 0i_{3}\,.
\end{equation}

Invoking Liebniz's rule for the application of covariant derivatives to
products of quaternions,
Eqs.\ (\ref{defETA},\ref{defZETA}) imply
\begin{equation}
\{ \eta , D\eta \} = 0\,,\;\;\{ \zeta , D\zeta \} = 0\,,\;\;\{ \xi , D\xi %
\} = 0\,.
\end{equation}
Therefore, we define
\begin{equation}
D_{\mu }\zeta  = {a}_{\mu }\eta  + {b}_{\mu }\zeta \eta\,,
\end{equation}
where ${a}_{\mu },\,{b}_{\mu }$ are real functions.

Then $D_{\mu }(D_{\nu }\zeta ) = {\cal A}_{\mu \nu } + \zeta {\cal B}%
_{\mu \nu }$, where
\begin{equation}
{\cal A}_{\mu \nu } = {a}_{\nu ,\mu }\eta - {a}_{\mu }%
{b}_{\nu }\,,\ %
{\cal B}_{\mu \nu } = {b}_{\nu ,\mu }\eta - {b}_{\mu }%
{b}_{\nu }\,.
\end{equation}

If ${a}_{\mu }$, (respectively ${b}_{\mu }$) vanishes, then so does
${\cal A}_{\mu \nu }$, (respectively ${\cal B}_{\mu \nu }$).
The ramifications of these circumstances will be explored below.

Now the additional postulate $D\eta \equiv 0$ implies for integer $k$,
\begin{equation}
\{\eta , D^{k}\zeta \} = 0\,,\ \{ \eta , D^{k}\xi \} = 0 .
\end{equation}
Hence, $D\zeta \propto \xi $ and $D\xi \propto \zeta$.  That is, the
compatibility condition is sufficient to force ${a} = 0$, decoupling the
symplectic components of the wavefunction in the absence of fully quaternionic
potentials.

Explicitly,
\begin{eqnarray}
\partial_{\mu }\eta &=& \theta_{,\mu}(\cos\theta \cos\phi i_{1} + \cos%
\theta \sin\phi i_{2} - \sin\theta  i_{3}) \nonumber \\
&& + \phi_{,\mu}(- \sin\theta \sin\phi i_{1} + \sin\theta \cos\phi i_{2})\,,
\end{eqnarray}
\begin{equation}
D_{\mu}\eta = \theta_{,\mu}\zeta + \phi_{,\mu}\sin\theta\xi + \case{1}{2}%
[{\cal Q}^{\eta}_{\mu}\eta + {\cal Q}^{\zeta}_{\mu}\zeta + {\cal Q}^{\xi}%
_{\mu}\xi , \eta] \equiv 0\,.
\end{equation}
Therefore ${\cal Q}^{\eta}_{\mu}$ is a free real parameter,
${\cal Q}^{\zeta}_{\mu} = \phi_{,\mu}\sin\theta $, and
${\cal Q}^{\xi}_{\mu} = -\theta_{,\mu}$.
Hence
\begin{eqnarray}
D_{\mu}\zeta &=& -\theta_{,\mu}\eta + \phi_{,\mu}\cos\theta \xi + \case{1}{2}%
[{\cal Q}^{\eta}_{\mu}\eta + {\cal Q}^{\xi}_{\mu}\xi , \zeta]\,, \nonumber \\
&=& ({\cal Q}^{\eta}_{\mu} + \phi_{,\mu}\cos\theta )\xi\,,
\end{eqnarray}
\begin{eqnarray}
D_{\mu}\xi  &=& \phi_{,\mu}(-\cos\phi i_{1} - \sin\phi i_{2}) + \case{1}{2}%
[{\cal Q}^{\eta}_{\mu}\eta + {\cal Q}^{\zeta}_{\mu}\zeta , \xi]\,, \nonumber \\
&=& -({\cal Q}^{\eta}_{\mu} + \phi_{,\mu}\cos\theta )\zeta\,,
\end{eqnarray}
so we have
\begin{equation}
{a}_{\mu} = 0\,,\;\;%
{b}_{\mu} = -({\cal Q}^{\eta}_{\mu} + \phi_{,\mu}\cos\theta )\,.
\end{equation}
We see that the imposition of the compatibility condition results in
the decoupling of the symplectic components of the quaternionic wavefunction.

\section{Q--Potentials}
In the nonrelativistic limit, we have the quaternionic analogue of the
Schr\"{o}dinger equation
\begin{equation}
(D_{t}\Psi)\eta + \case{1}{2m}{\bf D}\cdot {\bf D}\Psi = 0\,,
\end{equation}
equivalent to the pair of equations
\begin{mathletters}
\begin{equation}
(i\partial _{t} + \case{1}{2m}\Delta )\psi = 0\,,
\end{equation}
\begin{equation}
(i\partial _{t} + \case{1}{2m}\Delta )\varphi = ({b}_{t} + %
\case{1}{im}{\bf {b}}\cdot\bbox{\partial} - \case{1}{2m}{\cal B}^{k}_{\ k})%
\varphi\,.
\end{equation}
\end{mathletters}
That is, we carry out a local, quaternionic gauge transformation.
The $b$--field contains the remaining degrees of freedom associated with the
choice of a $\zeta$--field at each point of spacetime.
We do not consider here the physical meaning which might be ascribed to
restrictions on this procedure arising from the existence of topological
defects in the spacetime manifold.

These dynamical equations are independent of intrinsic spin,
and so an ensemble of spin states (correlated or statistical mixture) will
retain  its structure until a measurement of spin is made.  This would
imply that Bell experiments will be fundamentally unchanged by progressing
to QQM, but that the expectation values will be different to the CQM case
due to the new dependence of the Pauli spin operators on spacetime\cite{BJA}.

In the presence of a potential $V = V^{\eta} + \zeta V^{\zeta},\:V^{\zeta}%
\neq 0$, in the nonrelativistic limit, and, as in Eq.(\ref{GCovD}),
postulating the right-multiplication of the $\eta$--complex $V$--components on
the wavefunction, we have the pair of coupled dynamical equations
\begin{mathletters}
\begin{equation}
(i\partial _{t} + \case{1}{2m}\Delta - V^{\eta})\psi = -V^{\zeta}\varphi\,,
\end{equation}
\begin{equation}
(i\partial _{t} + \case{1}{2m}\Delta - V^{\eta} - {b}%
_{t} - \case{1}{im}{\bf {b}}\cdot\bbox{\partial} + %
\case{1}{2m}{\cal B}^{k}_{\ k})\varphi = V^{\zeta}\psi\,.
\end{equation}
\end{mathletters}
Note that this differs from \cite{BMAD} only by an irrelevant choice of
signs for the quaternionic components of the external potential.

Introducing Green functions ${\cal G}_{j}(x|x')$,
the most general solutions to these P.D.E.'s are
\begin{mathletters}
\begin{equation}
\psi = \psi_{o} - \int \!dx'{\cal G}_{1}(x|x')V^{\zeta}(x')\varphi(x')\,,
\end{equation}
\begin{equation}
\varphi = \varphi_{o} + \int \!dx'{\cal G}_{2}(x|x')V^{\zeta}(x')\psi(x')\,,
\end{equation}
\end{mathletters}
where $\psi_{o},\,\varphi_{o}$ are solutions to their respective homogeneous
P.D.E.'s, and we assume homogeneous boundary conditions.

Thus we have the formal, iterative solution for the symplectic components
\begin{eqnarray}
\psi &=& \psi_{o} - \int \!dx'{\cal G}_{1}(x|x')V^{\zeta}(x')%
\Bigl\{\varphi_{o}(x') \nonumber \\
&&+ \int \!dx''{\cal G}_{2}(x'|x'')V^{\zeta}(x'')\psi(x'')\Bigr\}\;\;%
\text{, etc.}
\end{eqnarray}

Hence, for small $V^{\zeta}$ we have
\begin{equation}
\psi = \psi_{o} - \int \!dx'{\cal G}_{1}(x|x')V^{\zeta}(x')%
\varphi_{o}(x') + {\rm O}(|V^{\zeta}|^{2})\,,
\end{equation}
and we can see that QQM with weak Q--potentials leads to inhomogeneous
dynamical equations.
Spontaneous creation and dematerialisation of particles is a consequence
of this situation.

\section{Interferometry Experiment}
To illustrate, we consider an interferometry experiment consisting of a
deBroglie--wave, say of  slow neutrons, split into two coherent beams.
One beam is passed through a constant, quaternionic potential of
bounded support, and then they are recombined to interfere \cite{AKSW}.

The final intensity pattern
\begin{equation}
{\cal I}(x,t) \propto \left|\psi  + [ S\psi - V^{\zeta}I_{[0,L]}%
(\varphi_{o}) + {\rm O}(|V^{\zeta }|^{2})]\right|^{2}\,,
\end{equation}
where $\psi $ is the reference wave, $|S|^{2}$ is the transmission coefficient
of the single barrier potential $V$ which is nonzero constant on the interval
$[0,L]$ and vanishes elsewhere, $\varphi_{o}$ is the time-dependent solution
of the corresponding hyper-complex component's homogeneous dynamical equation,
and
\begin{equation}
I_{[0,L]}(\varphi_{o}) = \int_{0}^{L}dx'{\cal G}_{1}(x|x')\varphi%
_{o}(x',t)\,.
\end{equation}

The Green function for the 1 dimensional, finite barrier potential is,
from \cite{MoFes},
\begin{equation}
{\cal G}_{1}(x|x') = -2\pi i(k_{1})^{-1}e^{ik_{1}|x-x'|}\,,
\end{equation}
where $k_{1} = (\omega _{1} - V^{\eta})^{1/2}$, and $i\partial _{t}\psi%
= \omega _{1}\psi$.
Assuming an effective constant potential, ${\cal V}_{Q}$, due to the presence
of Q--curvature in the interval $[0,L]$,  and that $i\partial _{t}\varphi%
_{o} = \omega _{2}\varphi$, we have in the region $x > L$,
\begin{eqnarray}
I_{[0,L]}(\varphi_{o}) &=& -2\pi i(k_{1})^{-1}{\cal N}_{\varphi}e^{%
ik_{1}x - i\omega _{2}t} \nonumber \\
&& \times \int_{0}^{L}\!\!dx'\bigl(e^{i(k_{2}-k_{1})x'} + R%
e^{-i(k_{2}+k_{1})x'}\bigr)\,, \nonumber \\
&&
\end{eqnarray}
where ${\cal N}_{\varphi}$ is a normalisation factor, $R>0$ due to reflection
within the barrier, and $k_{2} = (\omega _{2} - V^{\eta} - %
{\cal V}_{Q})^{1/2}$.
This is not translationally invariant, which is intuitively obvious as this is
a source problem.
Note that the condition for the intensity to be time independent is
that $\omega _{1} = \omega _{2}$.

In the case of a series of spatially bounded and nonintersecting
potential barriers, the order of traversal is critical (as
previously suggested\cite{BMAD,Peres}, but now for new reasons).
That is, given a set of weakly quaternionic potential barriers $V_{(n)}$
of width $L_{(n)}$, at positions $x_{(n)}$, to first order in the hyper-complex
barrier components we have as the QQM contribution to the particle
amplitude at the detector
\begin{equation}
-\sum_n V_{(n)}^{\zeta}I_{[x_{(n)},x_{(n)} + L_{(n)}]}(\varphi_{o}) +
{\rm O}(\{|V_{(n)}^{\zeta }|^2 \}).
\end{equation}
The Q--curvature will affect the $I$--integrals through the $\varphi_{o}$
field, and so a permutation of the order of the barriers will in general
result in a new intensity pattern being produced, which cannot be explained
merely in terms of a re-assignment of complex phases.

Thus, our quaternionic expectation value departs from the predictions of CQM
in a way qualitatively different to previous investigations of QQM in the
Q--flat limit.
In particular, QQM with nonvanishing Q--curvature manifests itself through
``external source'' effects, rather than through noncommuting phase factors.

\acknowledgements
One of the authors, S.P.B., acknowledges the support of an Australian
Postgraduate Research Award.  G.C.J.\ was supported by the Australian
Research Council and the University of Melbourne.

% REFERENCES


\begin{references}
\bibitem[*]{SPB}	e--mail address: spb@tauon.ph.unimelb.EDU.AU
\bibitem[\dag]{JOSHI}	e--mail address: joshi@tauon.ph.unimelb.EDU.AU
\bibitem{FJSS}  D.\ Finkelstein, J.\ M.\ Jauch, S.\ Schiminovich, and
		D.\ Speiser, J.\ Math.\ Phys.\ {\bf 3}, 207 (1962);
		{\bf 4}, 788 (1963).
\bibitem{Nash}	C.\ G.\ Nash and G.\ C.\ Joshi, Int.\ J.\ Theor.\ Phys.\
		{\bf 31}, 965 (1992);
		J.\ Math.\ Phys.\ {\bf 28}, 2883 (1987);
		{\bf 28}, 2886 (1987);
\bibitem{ARLH}	A.\ Razon and L.\ P.\ Horwitz, Acta Appl.\ Math.\
		{\bf 24}, 141 (1991); {\bf 24}, 179 (1991).
\bibitem{Hawk}  S.\ W.\ Hawking, Commun.\ Math.\ Phys.\ {\bf 87}, 395 (1982).
\bibitem{Adler}	S.\ L.\ Adler, Phys.\ Rev.\ D {\bf 37}, 3654 (1988);
		Phys.\ Rev.\ Lett.\ {\bf 57}, 167 (1986);
		Commun.\ Math.\ Phys.\ {\bf 104}, 611 (1986).
\bibitem{LHLB}	L.\ P.\ Horwitz and L.\ C.\ Biedenharn,
		Ann.\ Phys.\ {\bf 157}, 432 (1984).
\bibitem{BMAD}	A.\ J.\ Davies and B.\ H.\ J.\ McKellar,
		Phys.\ Rev.\ A {\bf 40}, 4209 (1989); {\bf 46}, 3671 (1992);
		A.\ J.\ Davies, Phys.\ Rev.\ D {\bf 41}, 2628 (1990).
\bibitem{Peres} A.\ Peres, Phys.\ Rev.\ Lett.\ {\bf 42}, 683 (1979).
\bibitem{algH}	The algebra of quaternions is the real span of a set of
		abstract units $\{1,\,i_1 ,\,i_2 ,\,i_3 \}$ with defining
		multiplication rules: $1i_j = i_j 1 = i_j,\ %
		i_j i_k = -\delta_{jk}1 + \sum_l \epsilon_{jkl} i_l$, %
		where $j,k,l = 1,2,3$.
\bibitem{BJwip} S.\ P.\ Brumby and G.\ C.\ Joshi, work in progress.
\bibitem{BJA}	S.\ P.\ Brumby, G.\ C.\ Joshi, and R.\ Anderson,
		University of Melbourne preprint UM--P--94/54 ; RCHEP--94/15.
\bibitem{AKSW}  A.\ G.\ Klein and S.\ A.\ Werner,
		Rep.\ Prog.\ Phys.\ {\bf 46}, 259 (1983).
\bibitem{MoFes} P.\ M.\ Morse and H.\ Feshbach, {\em Methods of Theoretical %
		 Physics} (McGraw--Hill, New York, 1953), Vol.\ 1.
\end{references}
\end{document}